\documentstyle[twocolumn,aps]{revtex}
\input epsf
\epsfverbosetrue

\begin{document}

\draft

\title{ A Pair of Interacting Spins Coupled to an Environmental Sea : \\
Dissipation and Mutual Coherence}

\author{M. Dub\'{e} and P.C.E. Stamp}

\address{ Physics Department, University of British Columbia, 6224 Agricultural
rd., \\  Vancouver, BC, Canada V6T 1Z1}

\maketitle

\begin{abstract}
We consider the quantum dynamics of two spin-1/2 systems, each coupled to a
bath
of oscillators, so that a bath-mediated coupling is generated between the
spins.
We find that the interaction destroys any coherent motion of the 2 spins, even
if the coupling of each spin to the bath is quite weak, unless the interaction
is extremely small. This is because the dynamic mutual bias between the spins
blocks any coherent transitions between nearly degenerate states. In many
quantum measurement operations this means that decoherence effects will be
much stronger during the actual measurement.

\end{abstract}

\pacs{PACS numbers:     }

In this paper we study the theoretical problem of a Pair of Interacting Spins
Coupled to an Environmental Sea (we use the acronym ``P.I.S.C.E.S.'' for
brevity). The 2 spins will each be described as 2-level systems, and we shall
concentrate on environments that can be represented using a Feynman-Vernon
oscillator bath \cite{feynman,leggett} model - at the end of the paper we
shall simply remark upon some analogous results obtained when the environment
can be represented as a spin bath \cite{prostamp1,prostamp2}. Our PISCES
problem
is clearly a generalisation of previous models like the ``spin-boson'' model,
describing a single spin coupled to a sea of oscillators - this model has been
widely studied in connection with the Kondo problem \cite{emery,tsvelick},
quantum diffusion \cite{prokagan}, SQUID tunneling \cite{leggett}, and
magnetic grains \cite{prostamp3,stamp}, and extensively reviewed (see, eg.,
refs. \cite{leggett,weiss}).

The interest of the PISCES model is in the way it reveals a dual role of the
environment - whilst it causes decoherence and dissipation in the motion of
each spin, one also expects that it might cause some mutual coherence between
them. For
very weak mutual coupling (as in the problem of two 2-level atoms coupled
radiatively) this mutual coherence is fairly well understood. However, as we
shall see, a radical change occurs once the coupling exceeds a very small
value, to be specified below. One then finds quantum relaxation behaviour of
the 2 spins, even at temperature $T=0$. Since the majority of PISCES
problems, in condensed matter and elsewhere, have couplings exceeding this
value, our results have important practical application. The PISCES model is
also a primitive model for quantum measurements, in which both dissipation and
the back-reaction of the measuring apparatus on the measured system are
included.

In cases where the environment is modelled by an oscillator bath, we will start
from an effective Hamiltonian $H_{PISCES} = H_{0} + H_{int}$, between spins
located at position ${\bf r_{1} }$ and ${\bf r_{2} }$, where
\begin{equation}
H_{0} = -\frac{\hbar}{2} ( \Delta_{1} \tau_{1}^{x} + \Delta_{2} \tau_{2}^{x} )
      + \frac{1}{2} \sum_{{\bf k}} m_{{\bf k}} (\dot{x}_{{\bf k}}^{2} +
\omega_{{\bf k}}^{2} x_{{\bf k}}^{2} )
\label{hbare}
\end{equation}

\begin{eqnarray}
H_{int} &=& \frac{1}{2} K_{ \alpha \beta } \, \tau_{1}^{\alpha } \,
\tau_{2}^{\beta } \nonumber \\
& & + \frac{1}{2} \sum_{{\bf k}} ( q_{01} c_{{\bf k}}^{(1)}
e^{i {\bf k} \cdot {\bf r_{1}}} \tau_{1}^{z} +
q_{02} c_{{\bf k}}^{(2)} e^{i {\bf k} \cdot {\bf r_{2}}} \tau_{2}^{z} ) x_{k}
\label{hint}
\end{eqnarray}
where $\tau^{\alpha}_{j}$ are the Pauli operators for each spin $(j=1,2)$,
and $ \{ x_{k} \}$ are the bath oscillators. In this paper we are really
interested in the indirect bath-mediated interaction between
$\mbox{\boldmath $\tau$}_{1}$ and $\mbox{\boldmath $\tau$}_{2}$, but we
include also a direct interaction $K_{\alpha \beta} ({\bf r})$, with
${\bf r} = {\bf r_{1}} - {\bf r_{2}}$; $K_{\alpha \beta} ({\bf r})$ can be
considered as modelling the static part of some other field-mediated
interaction \cite{note1}. We assume that ${\bf k}$ is a momentum index,
describing delocalised bath modes; we also ignore non-diagonal couplings
(ie., couplings in $\tau^{x}_{j}$ or $\tau^{y}_{j}$), which have a less
important effect on the physics \cite{leggett,prokagan}. Clearly our model
could be generalised, but at the expense of great complexity, and we believe
that (\ref{hbare}) and (\ref{hint}) bring out most of the essential physics.

The interactions with the bath are characterised by spectral functions
$J_{ij}( {\bf r}, \omega)$ whose diagonal elements $J_{j}(\omega ) \equiv
J_{jj}(\omega )$, referring to self-interactions of
$ \mbox{\boldmath $\tau$}_{j}$ via the bath, have the Caldeira-Leggett form
\cite{leggett}:
\begin{equation}
J_{j}(\omega ) = \frac{\pi }{2} \sum_{{\bf k}} \frac{ |c_{{\bf k}}^{(j)}|^{2}}
{m_{{\bf k}} \omega_{{\bf k}}} \delta( \omega - \omega_{{\bf k}} )
\label{sprecj}
\end{equation}

The off-diagonal elements have a complicated form if
$ | {\bf r} |$ is large, because of retardation effects \cite{note2}; these
begin to be important once $\omega_{{\bf k}} \geq v_{{\bf k}}/ | {\bf r} |$,
where $v_{{\bf k}}$ is the velocity of propagation of the $k^{th}$ oscillator
mode. Here, we shall ignore such effects (but see ref. \cite{dube}) which are
most important in the weak-coupling limit, so that
\begin{equation}
J_{12}({\bf r}, \omega ) = \frac{\pi }{2} \sum_{{\bf k}}
\frac{ |c_{{\bf k}}^{(1)} D_{{\bf k}}(0) c_{{\bf k}}^{(2) *} \, e^{i {\bf k}
\cdot {\bf r}}|} {m_{{\bf k}} \omega_{{\bf k}}} \delta( \omega -
\omega_{{\bf k}} )
\label{spec12}
\end{equation}
where $D_{{\bf k}}(0)$ is the static limit of the oscillator bath propagator
\cite{note2}.

We shall consider 2 forms \cite{note3} for $J_{ij}(({\bf r}, \omega )$.
The Ohmic form $J_{ij}({\bf r}, \omega ) = \omega \, \eta_{ij} \, e^{-\omega /
\Omega_{0}}$, with $\eta_{j} \equiv \eta_{jj}$ a local friction coefficient
and $\eta_{12} \sim (\eta_{1} \eta_{2})^{1/2} {\bf V}_{e}( {\bf r})$, is
appropriate to eg., electrons in Fermi liquid. The superohmic
$J_{ij}({\bf r}, \omega ) = \bar{g}_{ij}({\bf r}) \, \Theta_{D} \, ( \omega /
\Theta_{D})^{m} \, e^{-\omega / \Theta_{D}}$
applies to phonons or photons, with $m \geq 3$ in 3 dimensions \cite{prokagan};
now $\bar{g}_{12} \sim (\bar{g}_{1} \bar{g}_{2})^{1/2} {\bf V}_{\phi}
( {\bf r})$. For electrons $|{\bf V}_{e}({\bf r})| \sim \cos (2k_{F} r) /
(k_{F} r)^{3}$
(the RKKY form) in 3 dimensions; for phonons $|{\bf V}_{\phi}({\bf r})| \sim
(a_{0}/r)^{3}$, where $a_{0}$ is the lattice spacing. The values of the upper
cut-offs $\Omega_{0}$ and $\Theta_{D}$ are typically $\Omega_{0} \sim eV$ for
electrons, and $\Theta_{D} \sim 10^{-2} \, eV$ for phonons.

We also define a {\em mutual bias function} $\bar{\epsilon}({\bf r})$ as
\begin{equation}
\bar{\epsilon}({\bf r}) = - \frac{q_{01} q_{02}}{\pi} \int_{0}^{\infty}
    \frac{d \omega}{\omega} \, J_{12} ({\bf r}, \omega)
\label{bias}
\end{equation}
with sign such that positive/negative $\bar{\epsilon}({\bf r})$ corresponds to
ferromagnetic/antiferromagnetic coupling between $\mbox{\boldmath $\tau$}_{1}$
and $\mbox{\boldmath $\tau$}_{2}$.  Defining $\alpha_{ij}({\bf r}) =
\eta_{ij}({\bf r}) q_{0i} q_{0j} / 2 \pi \hbar$, one then has
\begin{equation}
\bar{\epsilon} ({\bf r}) = \left\{ \begin{array}{ll}
-2 \, \hbar \, \Omega_{0} \, \alpha_{12} ({\bf r}) & \ \ \ \ \mbox{(Ohmic)} \\
   -2 \, \hbar \, \Gamma (m) \, \Theta_{D} \, \bar{g}_{12} ({\bf r}) & \ \ \ \
     \mbox{(Superohmic)}
                            \end{array} \right.
\label{biasform}
\end{equation}
with $\bar{\epsilon}$ proportional to the upper cut-off frequency in both cases
 \cite{note1}.

A complete study of this problem calculates the 16 elements of the
2-spin density matrix $\hat{\rho} \,
(\mbox{\boldmath $\tau$}_{1},\mbox{\boldmath $\tau$}_{2}, t;
\mbox{\boldmath $\tau$}_{1}',\mbox{\boldmath $\tau$}_{2}', t')$.
To do this we use standard influence functional techniques
\cite{feynman,leggett}; the total influence functional
$F \{ \mbox{\boldmath $\tau$}_{1}, \mbox{\boldmath $\tau$}_{2} \}$
can be factorised as $F = F_{11} \, F_{22} \, F_{12}$, where $F_{11}$ and
$F_{22}$ are identical to the
single spin-boson functionals, and $F_{12}$ contains all dynamic couplings via
the bath. $F_{12}$ is a functional over the paths of both spins together, and
is thus intrinsically more complex than $F_{11}$ and $F_{22}$; it cannot be
evaluated in closed form for arbitrary values of the parameter in (\ref{hint}).
The ``dilute-blip'' approximation \cite{leggett} cannot be applied to
$F_{12}$ in the very weak coupling limit because then one may have overlap in
time between arbitrary configurations of each spin - thus this limit must be
handled perturbatively \cite{dube}. Away from this limit, the mutual bias
$\bar{\epsilon}({\bf r})$ plays an essential role. As a function of time
the coupling energy in a path integral contribution to $\hat{\rho}(t)$ flips
between $\pm \bar{\epsilon}({\bf r})$; once $|\bar{\epsilon}({\bf r})|$ is
large enough its effect is to suppress configurations in which both spins
are simultaneously in blip states. One may then use a dilute-blip
approximation for $F_{12}$. The parameter regime in which this works depends
on the precise form of $J_{ij}({\bf r},\omega )$; we give the limits of
validity below, for the ohmic and superohmic cases \cite{fisher}

There is no space here to give results for all matrix elements of $\hat{\rho}$
(see ref. \cite{dube}); here we concentrate on $P_{++}(t)$, the probability to
find the system in the state $| \uparrow \uparrow \rangle$ at time $t$, if at
$t=0$ it was also in state  $| \uparrow \uparrow \rangle$. Then for
ferromagnetic coupling ($\alpha_{12} > 0$) we find
\begin{equation}
P_{++}(t) = A_{-} + A_{+} e^{ - ( \Gamma_{1} + \Gamma_{2} )t }
        + B_{+} e^{-R_{+}t} + B_{-} e^{-R_{-}t}
\label{ppp}
\end{equation}
\begin{equation}
A_{\pm} = \frac{1}{4} [1 \pm \tanh(\bar{\epsilon} /2kT)]
\label{constA}
\end{equation}
\begin{equation}
B_{\pm} =  \frac{1}{4} \left[ 1 \mp  \frac{\Gamma_{1}+\Gamma_{2}}{\sqrt{R}}
 \tanh(\bar{\epsilon} /2kT) \right]
\end{equation}
\begin{equation}
R_{\pm} = \frac{1}{2}(\Gamma_{1}+\Gamma_{2}) \pm \frac{1}{2} \sqrt{R_{12}}
\end{equation}
\begin{equation}
R_{12}(T) = \Gamma_{1}^{2} + \Gamma_{2}^{2} -2\Gamma_{1}\Gamma_{2} \left[ 1-2
\tanh^{2} (\bar{ \epsilon} /2kT) \right]
\end{equation}
with the rates $\Gamma_{1}$ and $\Gamma_{2}$ being those for a spin-boson
system in a bias $\bar{\epsilon}$ \cite{leggett} :

\begin{equation}
\Gamma_{j} = \Delta_{j}^{2} \int_{0}^{\infty} dt \,  e^{ - Q_{2}^{(j)}(t) }
\, \cos \, [Q_{1}^{(j)}(t) \, ] \, \cos \, [ \bar{\epsilon} t/ \hbar ]
\end{equation}

\begin{equation}
Q_{1}^{(j)}(t) = \frac{q_{0j}^{2}}{\pi\hbar}  \int_{0}^{\infty} d\omega \,
\frac{J_{jj}(\omega )}{\omega^{2}} \, \sin \omega t \, .
\end{equation}

\begin{equation}
Q_{2}^{(j)}(t) \equiv \frac{q_{0j}^{2}}{\pi\hbar}  \int_{0}^{\infty} d\omega \,
\frac{J_{jj}(\omega )}{\omega^{2}} \,[1-\cos\omega t]\coth(\beta\hbar\omega/2)
\end{equation}

In this way the complete solution is seen to be one involving quantum
relaxation of $\hat{\rho} (t)$, with the relaxation rates $( \Gamma_{1} +
\Gamma_{2})$ and $R_{\pm}$ depending essentially on $\bar{\epsilon} / kT$. To
see how this works in practice, consider first the Ohmic case, where
\cite{leggett}
\begin{equation}
\Gamma_{j} = \frac{\Delta_{j}^{2}}{2\Omega_{0}}
\left[ \frac{2\pi kT}{\hbar\Omega_{0}} \right]^{2\alpha_{j}-1}
\frac{\cosh(\bar{\epsilon}/2kT)}{\Gamma(2\alpha_{j})}
| \Gamma(\alpha_{j}+i\bar{\epsilon}/2\pi kT) |^{2}
\end{equation}
with the functions $\Gamma(x)$ on the right-hand side being Gamma functions.
If there were no interactions at all between $ \mbox{\boldmath $\tau$}_{1}$ and
$ \mbox{\boldmath $\tau$}_{2}$, then at $T=0$, the
$ \mbox{\boldmath $\tau$}_{j}$ would
freeze for $\alpha_{j} > 1$, relax exponentially (with power law corrections)
for $ 1/2 < \alpha_{j} < 1$, and show damped oscillations for $\alpha < 1/2$;
at any finite $T$, the $ \mbox{\boldmath $\tau$}_{j}$ would relax exponentially
for any value of the $\alpha_{j}$ (see ref. \cite{leggett}).

Switching on the interaction $J_{12}({\bf r},\omega)$ quickly changes this.
Suppose, as above, that $\alpha_{12}({\bf r})$ is positive (ferromagnetic).
Then once $ \bar{\epsilon} \gg kT$, ie., once $\alpha_{12} \gg kT/ \Omega_{0}$,
so that $\tanh (\, \bar{ \epsilon} /2kT \, ) \sim -1$, then $P_{++}(t) = 1$
for all times of interests, ie., the 2 spins freeze each other ! More generally
we see that at long times $P_{++}(t)$ relaxes to $\frac{1}{4}
[1+\tanh(\alpha_{12} \, \Omega_{0}/kT)]$. A similar freezing to the
antiferromagnetic $| \uparrow \downarrow \rangle$ occurs if
$ - \alpha_{12} \gg kT/ \Omega_{0}$. Since $\Omega_{0}$ is usually large,
this freezing can happen for very small
coupling at low $T$. The results are valid if $\Delta_{j} \ll | \alpha_{12} |
 \Omega_{0}$, at any $T$, or for any $\alpha_{12}$ if $kT \gg \Delta_{j} /
 \alpha_{j}$. In cases where the $\Delta_{j}$ describe tunneling matrix
elements, as in SQUID's \cite{leggett}, quantum diffusion \cite{prokagan},
or magnetic grains \cite{stamp}, the tunneling rates are often rather small
and so the results apply for very small couplings $\alpha_{12}$; the same
will be true for many metallic glasses.

Similar results are obtained in the superohmic case, for which
\cite{leggett,prokagan}
\begin{equation}
\Gamma_{j} = ( q_{0j} \Delta_{j} / \bar{\epsilon} )^{2} \,
J_{j} (\bar{\epsilon}) \, \coth ( \bar{\epsilon} / 2kT)
\end{equation}
In the common case where $m=3$, this gives $\Gamma_{j} \sim q_{0j}^{2}  \,
 \bar{g}_{12}^{2} \, \Gamma (3) \, \Delta_{j}^{2} / \, \Theta_{D}$. If
$ |\bar{g}_{12}| \, \Theta_{D} \gg kT$, the spins will again be frozen;
otherwise
we get relaxation according to (\ref{ppp}), provided $\Delta_{j} \ll
| \bar{g}_{12}| \, \Theta_{D}$, for any $T$, or for any $\bar{g}_{12}$ if
$ kT \gg \Delta_{j} / \bar{g}_{j} \, \Theta_{D}$. Typically, $\Delta_{j} \ll
\Theta_{D}$ and this relaxation will be rather slow. This result shows that the
bias $\bar{\epsilon}$ plays a crucial role - if there were no coupling between
the spins, then they would not freeze even at $T=0$, no matter how large the
couplings $\bar{g}_{1}$ and $\bar{g}_{2}$.

Some qualitative features of the results deserve special mention; in
particular

(i) It is quite common in phenomenological studies of this problem to simulate
the environmental effects on the 2 coupled spins by stochastic terms
\cite{forster}. In general such methods will give results quite unlike these
(often with oscillatory behaviour in $\hat{\rho}(t)$ for large inter-spin
couplings). The reason for the discrepancy is fundamentally that quantum
influence functionals cannot be modeled by classical (c-number) noise source
\cite{feynhibbs}.

\begin{figure}[ht]
\epsfysize=3.0in
\epsfbox[-0 132 575 700]{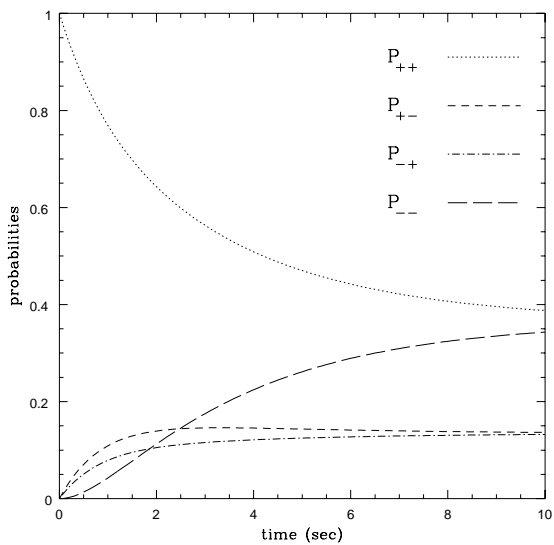}
\caption{Probabilities of occupation for a system with parameters
$\Gamma_{1} = 0.5 \, \mbox{sec}^{-1}$, $\Gamma_{2} =0.7 \, \mbox{sec}^{-2}$
and $\bar{\epsilon}/2 k T = -0.5$. The system starts at $t=0$ in the state
$ | \uparrow \uparrow \rangle$; $P_{\alpha \beta}$ is then the probability
that at time $t$ the system has $\tau_{1}^{z} = \alpha$ and $\tau_{2}^{z} =
\beta$.}
\label{fig1}
\end{figure}

(ii) The 3 relaxation times appearing in $\hat{\rho}(t)$ (see Eq.(\ref{ppp}))
can be quite different, and their contributions will in general appear in
different proportions in the 16 different elements of $\hat{\rho}(t)$.
Fig.(\ref{fig1}) illustrates this by plotting 4 of these matrix elements
(including $P_{++}(t)$) calculated in the same way as (\ref{ppp}).

(iii) There is no trace left in our results of the ``phase-transition''
occuring at $\alpha = 1/2$ and $\alpha =1$ in the spin-boson problem at $T=0$,
for the case of ohmic dissipation \cite{leggett}. The localisation transition
at $\alpha =1$ is replaced here by a crossover to localised behaviour, at
$T=0$, as $ | \bar{\epsilon}({\bf r}) | = 2 \, \Omega_{0} \, \alpha_{12}
({\bf r})$ becomes greater than the larger of $\Delta_{1}$ and $\Delta_{2}$.
We can understand this physically as ``degeneracy blocking'', ie., the mutual
bias blocks tunneling by destroying the degeneracy between the initial and
final states. There is also a finite temperature crossover to localisation,
for a finite coupling $\bar{\epsilon}({\bf r}) \gg \Delta_{j}$, as one lowers
$T$; this occurs for $ kT \sim 2 \, \Omega_{0} \, \alpha_{12}  ({\bf r})$,
for ferromagnetic coupling, and $ kT \sim -2 \, \Omega_{0} \, \alpha_{12}
({\bf r})$, for antiferromagnetic coupling. Analogous crossovers will also
exist for superohmic coupling.

A surprising feature of the results (at least for us) is that our solution
(\ref{ppp}) does not allow any mutually coherent oscillations of the two spins.
Thus, eg., in the ohmic problem, if $\alpha_{1},\alpha_{2} < 1/2$, so each spin
exhibits damped oscillations if $\alpha_{12} = 0$, one might have expected to
find that if $\Delta_{1} \sim \Delta_{2}$, switching on $\alpha_{12}$ could
cause some kind of ``frequency locking'' between the 2 spins. Instead one
simply finds that the mutual bias destroys the oscillations, once we are out
of the very weak perturbative limit. A corollary of this is that if we start
off with the 2 coupled spins and no bath, then the effect of coupling in the
bath is to destroy coherent oscillations of each spin far more quickly than one
would expect from the known destructive effects of the bath on single spin
coherence \cite{leggett}.

We have no space to discuss detailed applications of these results to
particular
systems (particularly as $J_{12}({\bf r},\omega )$ is often long range in
${\bf r}$, so that for a system containing many 2-level systems, it is not
sufficient to simply examine the 2-spin correlation functions). However our
main result, that the coherence-destroying role of the bath is greatly
exaggerated once mutual coupling between 2 (or more) systems is allowed, seems
to be of general relevance. In particular it shows that during a quantum
measurement, decoherence effects will be much stronger {\em during the
measurement operation}. Consider, eg., a SQUID exhibiting damped coherent
oscillations \cite{leggett} for $\alpha < 1/2$. It is clear that when this
SQUID is coupled to a measuring apparatus (eg., another SQUID \cite{tesche},
modelled also as a damped oscillatory 2-level system), the coherence will be
destroyed unless the mutual coupling $\alpha_{12}$ is extremely small, ie.,
coherence will only persist whilst system and the apparatus are decoupled.
A detailed analysis of this example \cite{dube} takes account of the
coupling of each SQUID to the same set of EM field modes (whose detailed form
will depend on sample geometry). The inductive part of the interaction which
is responsible for the measurement is {\em generated}, as in the theory of this
paper, by this coupling to the EM bath. One must also account for the resistive
coupling of each SQUID's flux coordinate to separate quasiparticle baths.

Finally we note that application of our results to many real systems will be
complicated by the effects of the ``spin-bath'' of nuclear spins and
paramagnetic impurities \cite{prostamp1,prostamp2}, since these spins (a)
add a further bias field acting on $\mbox{\boldmath $\tau$}_{1}$ and
$\mbox{\boldmath $\tau$}_{2}$, and (b) can be flipped by a transition of
$\mbox{\boldmath $\tau$}_{2}$ or $\mbox{\boldmath $\tau$}_{2}$, thereby
causing further decoherence. The bias field depends on the polarisation of the
nuclear spins, and is particularly important if the system is magnetic. However
it can even play a role in SQUID's, because although the SQUID flux couples
only
very weakly to the nuclear spins, it couples to a lot of them, and the total
energy can easily exceed $\Delta_{j}$ by several orders of magnitudes.

This work was supported by NSERC in Canada. We are also greatful to R.F. Kiefl,
N.V. Prokofev, and A. Wurger for useful comments.

\newpage



\begin{references}
\bibitem{note2}  Retardation effects will be more important for phonons than
for electrons at low T, since $\Theta_{D} \ll \Omega_{0}$. We emphasize here
that our formalism is not altered by retardation (we simply replace
$D_{{\bf k}}(0)$ in (4) by the full propagator $D_{{\bf k}}( \omega )$),
but $J_{12}( {\bf r}, \omega)$ can no longer be factorised in
general, so the results are more complex - see ref. 13 for details.
\bibitem{note3} One can formally consider the ``subohmic'' form $J(\omega )
\sim \omega^{s}$, with $s < 1$ (cf. ref. [2]); but then $\bar{\epsilon}$
diverges in the infrared, and the model is physically meaningless
\end{references}
\end{document}